\documentclass[aps,pra,twocolumn,showpacs,superscriptaddress]{revtex4}  
\usepackage{graphicx}  
\usepackage{dcolumn}   
\usepackage{bm}        
\usepackage{amssymb}   
\usepackage{verbatim}  
\usepackage{epstopdf}
\usepackage{color}

\newcommand{\beq}{\begin{equation}}
\newcommand{\eeq}{\end{equation}}

\newcommand{\njp}{New. J. Phys.}

\begin{document}

\title{Correlated bosons in a one-dimensional optical lattice: Effects of the trapping potential and of quasiperiodic disorder}

\author{Uttam Shrestha}
\affiliation{LENS, Universit\`a di Firenze, via Nello Carrara 1, 50019 Sesto Fiorentino, Firenze, Italy}
\author{Michele Modugno}
\affiliation{LENS, Universit\`a di Firenze, via Nello Carrara 1, 50019 Sesto Fiorentino, Firenze, Italy}
\affiliation{Dipartimento di Fisica e Astronomia, Universit\`a di Firenze,  50019 Sesto Fiorentino,  Italy}
\affiliation{Department of Theoretical Physics and History of Science, Universidad del Pa\'is Vasco UPV-EHU, 48080 Bilbao, Spain}
\affiliation{IKERBASQUE, Basque Foundation for Science, 48011 Bilbao, Spain}

\date{\today}

\begin{abstract}

We investigate the effect of the trapping potential on the quantum phases of strongly correlated ultracold bosons in one-dimensional periodic and quasiperiodic optical lattices. By means of a decoupling meanfield approach, we characterize the ground state of the system and its behavior under variation of the harmonic trapping, as a function of the total number of atoms. For a small atom number the system shows an incompressible Mott-insulating phase, as the size of the cloud remains unaffected when the trapping potential is varied. When the quasiperiodic potential is added the system develops a metastable-disordered phase which is neither compressible nor Mott insulating. This state is characteristic of quasidisorder in the presence of a strong trapping potential. 

\end{abstract}

\maketitle 

\section{\label{sec:level1}Introduction}

The study of quantum phases for the strongly correlated system of identical particles in a lattice is a core research topic in condensed matter and solid state physics~\cite{Sachdev}. Due to rapid advances in the field of ultracold atomic physics such a system has been routinely realized in optical lattice experiments, mimicking condensed matter physics with unprecedented precision~\cite{BlochRMP}. Indeed, ultracold atoms in optical lattices are considered as promising quantum simulator for the strongly correlated electron system in solid state physics~\cite{Jakcsh}. Some of the 
hallmarks of such trapped quantum gases in a lattice are the experimental realization of the superfluid-Mott insulator transition~\cite{Greiner}, of a Tonks–Girardeau gas~\cite{Paredes}, of the Anderson localization for a Bose-Einstein condensate with tunable interactions~\cite{Raoti}, and many others.

In the experiments ultracold systems are usually realized in the presence of external parabolic confinement, hereafter, we synonymously call it \emph{trapping potential}. This makes the ultracold atomic system spatially local since the local chemical potential varies at each spatial point due to inhomogeneity. The competition between the atom-atom interaction energy and the local confining potential at each point dictates the true state of the system. In particular, in a strongly correlated system with random or quasiperiodic disorder the trapping potential may have a dramatic effect in the ground state properties and thus demands a detail and systematic investigation. In fact, there have been a number of studies that show the importance of the inhomogeneity in the lattice system~\cite{Batrouni}.

A Mott state for the trapped fermionic atoms in an optical lattice and its incompressible nature has recently been studied experimentally~\cite{Jordens, Schneider}. The system's negligible response with the variation of the trapping potential is the characteristic feature of the incompressible Mott insulating state. A similar line of thought has also been proposed in the context of detecting Bose-glass phase in a lattice system with random or quasi-random lattice potential~\cite{Delande,Roscilde,RoscildeNJP}. This technique the so-called the trap-squeezing spectroscopy (TSS) relies on the response of the system with the variation of the trapping potential, since the size~\cite{Delande} and the central density~\cite{RoscildeNJP} of the system depend on the strength of the trapping potential.

In Ref.~\cite{Delande} it is reported the fact that the variation of the radius of atomic cloud with the confining potential may be a useful tool for detecting the disordered phase such as the Bose-glass~\cite{Fisher}. It should be noted that for a trapped system there is a nonvanishing superfluid fraction since the atoms near the edges can be delocalized with significant number fluctuations. In this sense it is more precise to call the superfluid (Mott) phase in a trapped system as the phase with high (low) degree of compressibility. Finding a true signature of incompressible region over many lattice sites requires a strong lattice with minimal edge fluctuations. The Mott and the superfluid region can be dramatically affected by the variation of the external trapping potential as well as the total number of atoms present in the system since the chemical potential depends on both of them. Fixing the chemical potential by varying the atom number when the trap frequency is changed has also been proposed in~\cite{Roscilde}. However, in real experimental situation this restriction poses two challenges: (i) the chemical potential is not directly accessible in an experiment, and (ii) the number of atoms in each experimental shot is fixed. 
 
In this paper we follow a more direct approach. We study the properties of strongly interacting bosonic atoms in an optical lattice in the presence of external trapping and additional quasiperiodic potential. For a strong lattice the tunneling is exponentially suppressed so that the system can be in deep Mott regime with vanishing number fluctuation. We first develop a phase diagram in the parameter space spanned by the total number of atoms and the external trapping potential. In a pure system the phase diagram consists of alternate stripes of incompressible Mott and compressible superfluid region. Such compressible-incompressible region can also be obtained by studying the variation of the root mean square (RMS) size of the cloud with the number of atoms as well as with the trapping potential. In the Mott state the size of the cloud varies linearly with the atom number since the available state for additional particle is located at the edge of the cloud. For a small number of atoms the incompressible Mott state is robust with respect to the variation of the trapping potential. This state becomes compressible when the disorder is introduced by means of an additional lattice potential. In the quasiperiodic system with the fixed number of atoms, the decrease in cloud size with increase in trapping potential along with the vanishing superfluid fraction suggests that the system is in compressible Bose-glass phase. In addition to the compressible state, the RMS size of the cloud also develops plateaus as the trapping potential becomes stronger indicating the presence of incompressibility in the disordered system. This state differs from the Mott phase particularly in their density distribution, and is  characteristics of the quasiperiodic potential in the presence of trapping potential. Similar findings but in different model in the presence of off-diagonal disorder has also been reported using quantum Monte-Carlo simulations~\cite{Sengupta}. 

The paper is organized as follows. In Sec.~II we introduce the Bose-Hubbard model in a one-dimensional lattice in the presence of additional modulating potential. We employ decoupling mean-field method (DMF)~\cite{Sheshadri} to solve the ground state of the Hamiltonian. For completeness, we also connect the Bose-Hubbard parameters approximately to the experimentally measurable quantities such as lattice heights, atomic scattering lengths etc. In Sec.~III we study the properties of the system particularly the quantum phases as the trapping potential and the atom number are varied. Finally in Sec.~IV we study the effect of quasiperiodicity, in particular, the possible modifications in quantum phases and the appearance of disordered Mott-like incompressible phase.

\section{\label{sec:level1}Quasi Disorder Bose-Hubbard Model}

At zero temperature, the quantum state of bosonic atoms in a one-dimensional deep optical lattice with diagonal disorder is approximately governed by the modified Bose-Hubbard (BH) Hamiltonian~\cite{Fisher,Jaksch98}
\beq
\hat H=\sum_{i=1}^{L}\Bigg{[}-J(\hat{a}^\dag_i \hat{a}_{i+1}+\text{h.c.})+\frac{U}{2}\hat n_i(\hat n_i-1)-\mu \hat n_i+\epsilon_i \hat n_i\Bigg{]}\, ,
\label{Eq1}
\eeq
where $\hat{a}^\dag_i $ and $\hat{a}_i $ are the usual creation and annihilation operators for a boson at the $i^{th}$ lattice site, and $\hat n_i=\hat{a}^\dag_i  \hat{a}_i$ is the corresponding number operator. 
Here, $J$ is the kinetic energy for an atom hopping to the nearest site, $U$ is the on-site two-body interaction energy, $\mu$ is the global chemical potential, and $\epsilon_{i}$ is the on-site energy of an atom. The latter is composed of both harmonic potential resulting from magnetic and optical fields and disordered potential generated by a secondary lattice~\cite{Aubry,Fallani},
\beq
\epsilon_i=\frac{1}{2} {\cal V}_{t} {z_i}^2+\Delta\cos^2(\pi\alpha i+\delta)\, .
\eeq
Here the trapping strength is ${\cal V}_{t}=m{\omega_z}^2d^{2}$, $m$ being the mass of an atom, $\omega_z$ the longitudinal trapping frequency and $d$ the lattice spacing. The incommensurate parameter $\alpha$ is the ratio between the wavelengths of the primary and secondary laser lights, whereas $\Delta$, which is proportional to the intensity of the secondary lattice, measures the strength of the disorder~\cite{Modugno}. The relative phase between the primary and the secondary lattices $\delta$ is chosen in such a way that the harmonic potential minimum lies on one of the minima of the secondary lattice potential, preferably chosen at the central lattice site. All the parameters discussed in this model can be expressed in terms of experimentally measurable quantities such as the heights of the primary and secondary lattices, the wavelengths of the lasers, the s-wave scattering length characterizing the atom-atom interactions etc.~\cite{Modugno,Zwerger} (see Sec.~III).

In a homogeneous case the Hamiltonian (\ref{Eq1}) shows the superfluid-Mott insulator transition when the ratio $J/U$ exceeds some critical value $(J/U)_c$~\cite{Greiner,Fisher}. In a weakly interacting regime, the tunneling term in the Hamiltonian dominates over interactions, and the corresponding many-particle ground state is the superposition of all single particle wave functions spread over the entire lattice, causing the system to be superfluid. On the other hand, if the tunneling term is negligible the ground state energy of the interacting system is minimized  with a well defined number of atoms per site and the many-body ground state is a product of Fock states at each lattice site leading to the Mott insulating state. In this regime the spectrum presents a gap of the order of $U$, corresponding to the displacement of an atom to an occupied site.

In addition to these phases the BH model also predicts an insulating but gapless phase in the presence of random disorder, known as Bose-glass~\cite{Fisher}. However, the search and detection of this phase has been a long standing experimental challenge from the time of its conception, and so far there are preliminary experimental indications~\cite{Fallani} but no clear evidence that directly supports its existence.

The quantum properties of lattice bosons governed by the BH model can be studied accurately through exact numerical diagonalization only for a few lattice sites and a small number of particles~\cite{Roth}. Other methods, such as quantum Monte Carlo (QMC) simulations~\cite {Roscilde,Wessel}, density matrix renormalization group (DMRG)~\cite{Roux} and  time-evolving block decimation (TEBD)~\cite{Vidal} have been extensively used to extract the information on the ground state and excitation properties of the system. In this paper we use the decoupling mean-field (DMF) approximation~\cite{Sheshadri}, that provides a reasonable picture of the phase diagram of the homogeneous system for both pure and disordered case. This method also parallels to another popular mean-field approach the so-called the Gutzwiller approximation~\cite{Buosante}. 

In the DMF approach one can approximate the off-diagonal terms in the Hamiltonian as
\beq
\hat{a}^\dag_i \hat{a}_{i+1}\approx \langle \hat{a}^\dag_i\rangle \hat{a}_{i+1}+\hat{a}^\dag_i \langle\hat{a}_{i+1}\rangle - \langle\hat{a}^\dag_i\rangle\langle \hat{a}_{i+1}\rangle\, ,
\eeq
yielding the mean-field Hamiltonian 
\begin{widetext}
\beq
\hat{H}_{MF}=\sum_{i=1}^{L}\Bigg{[}-J({\phi}^*_i \hat{a}_{i+1}+\hat{a}^\dag_i \phi_{i+1} - {\phi}^*_i{\phi}_{i+1} +\text{h.c.})
+\frac{U}{2} \hat n_i(\hat n_i-1)-\mu \hat n_i+\epsilon_i \hat n_i\Bigg{]}\, ,
\eeq
\end{widetext}
where $\phi_i\equiv \langle \hat{a}_i\rangle $ is the superfluid order parameter, that provides a residual coupling between neighboring lattice sites. By using periodic boundary conditions the Hamiltonian can be written as the sum of the on-site hamiltonians as $\hat{H}_{MF}=\sum_i \hat{H}_i$ with
\begin{widetext}
\beq
\hat{H}_i=J(\phi_{i-1}^*+\phi_{i+1}^*)\hat{a}_{i}+J(\phi_{i-1}+\phi_{i+1})\hat{a}_{i}^\dag
\nonumber-J(\phi_{i-1}+\phi_{i+1})\phi_i
+\frac{U}{2}\hat n_i(\hat n_i-1)-\mu \hat n_i+\epsilon_i \hat n_i\,.
\eeq
\end{widetext}
The Hamiltonian can be diagonalized self-consistently using the number basis and the ground state can be found for a given set of parameters $J,\,U,\text{\,and\,}\Delta$~\cite{Sheshadri}.

\subsection{System geometry and parameters}

In recent experiments it has been considered a system of ultracold bosons in the presence of two strong transverse optical lattices of intensity $s_\perp$  in units of photon recoil energy, creating a two dimensional array of independent tubes~\cite{Paredes}. For large values of $s_\perp$, the coupling between different tubes vanishes, so that each tube portrays a quasi 1D atomic system. 
 
In order to study the feature governed by the Hamiltonian~(\ref{Eq1}) a lattice can be created
in the tube direction ($z$-axis) by shining a third pair of counter-propagating laser of wavelength $\lambda_1$ of lower intensity, $s_{1}^z\ll s_\perp$. Quasidisorder can be incorporated by introducing an additional pair laser of much smaller intensity $s_{2}^z$ and of wavelength $\lambda_2$ incommensurate with the main lattice, producing an effective quasiperiodic potential~\cite{Fallani}. Such a controlled disorder essentially captures the features of random disorder as demonstrated in a recent experiment on Anderson localization~\cite{Raoti}, and is therefore a suitable medium to study disorder induced quantum phases such as Bose-glass. 

Under these conditions, the BH parameters can be tuned with the experimental measurable quantities, according to~\cite{Modugno, Gerbier}
\begin{eqnarray}
J &\simeq& \frac{1.43}{\sqrt\pi} ({s_{1}^z})^{0.98} \exp(-2.07\sqrt{s_{1}^z})\, ,
\\
U &\simeq& 5.97(a_s/\lambda_1)(s_1^z~s_\perp ^2)^{0.88/3}\, ,
\\
\Delta&\simeq&\frac{s_2^z\beta^2}{2}e^{\displaystyle-\beta^\alpha/({s_1^z})^{\gamma}}\, ,
\label{eq:delta}
\end{eqnarray}
in terms of the recoil energy $E_R=2\hbar^2\pi^2/m\lambda^2$. Here $a_s$ is the s-wave scattering length and $\alpha,\,\beta ~\rm{and}~ \gamma$ are fitting parameters. Each laser beam also creates an additional harmonic confinement (due to the focusing) that introduces a source of inhomogeneity; therefore the number of atoms in each tube varies with maximum density at the central tube. In the experiments these numbers also depend upon the loading procedure, the adiabaticity of the lattice ramping and the external harmonic confinement. If the cloud follows the Thomas-Fermi (TF) profile, one can write the atom distribution in each tube as~\cite{Kramer}
\beq
N_{x,y}=N_{0,0}\bigg(1-\frac{x^2+y^2}{R_{TF}^2}\bigg)^{3/2}\, ,
\eeq
where $(x,y)$ represents the position of the tube in the transverse plane, $R_{TF}$ is the TF radius given by
\beq
R_{TF}=\sqrt{\frac{\hbar \bar \omega}{m \omega_\perp ^2 d^2}} \bigg[\frac{15 N a_s d^2}{a_{ho}} \bigg(\int dx {f_0}^4 (x)\bigg)^2 \bigg]^{1/5}\, ,
\eeq
with $N$ being the total number of atoms, and $N_{0,0}=5N/(2 \pi R_{TF}^2)$ the number of atoms in the central tube.
Here $m$, $a_s$ and $d$ are respectively the mass, the s-wave scattering length and the lattice spacing in the direction of the tube, whereas $\omega_\perp$ and $\bar \omega=(\omega_\perp^2 \omega_z)^{1/3}$ are the transverse and averaged trapping frequencies, $a_{\rm ho}=\sqrt{\hbar/m\omega_z}$ is the harmonic oscillator length, and $f_0$ is a function localized at the origin. 

For the discussion presented in this paper we choose the typical parameters of the experiment in Ref.~\cite{Fallani}. In particular, for a system of $N=2\times 10^5$  $^{87}$Rb atoms with s-wave scattering length $a_s=100~{\rm a}_{0}$, in trapping frequencies $\omega_{\perp}=2\pi\times90$ Hz and $\omega_{z}=2\pi\times77$ Hz, and the transverse lattice height $s_{\perp}=40$, the number of atoms in the central tube and the average number of atoms in each tube are respectively $N_{0,0}\approx 200$ and $N_{av}\approx 80$.

Since our model is strictly one-dimensional we fix the relevant parameters as stated in the previous paragraph by paying attention to a single tube only. In all our calculations the height of the primary lattice is fixed to  $s_{1}^z=16$ corresponding to the ratio $U/J\simeq 132$. The length of the lattice is fixed to $L=64$. We study the properties of the system by varying both the number of atoms in a given tube and the trapping potential. 
 
\section{\label{sec:level1}Primary lattice only: Pure case}

In this section we analyze the behavior of the system with the variation of the confining potential and the total number of atoms for the case when there is no additional secondary lattice. The state of the system can be characterized by considering several quantities, including  condensate fraction, superfluid fraction, number fluctuations, defined as follows.

The \textit{condensate fraction} is defined as the ratio between the number of condensate atoms to the total number of atoms, and can be simply expressed in terms of the order parameter $\phi_i$ as 
\beq
f_c= \frac{\sum_i^L|\phi_i|^2}{{\cal N}_{t}}\,,
\eeq
where ${\cal N}_{t}$ is the number of atoms in the tube.
The \textit{superfluid fraction}, on the other hand, can be calculated by measuring the response of the system to an infinitesimal flow, by imposing twisted boundary conditions~\cite{Roth,Burnett} that corresponds to a phase gradient across each lattice site. The superfluid component responds to the velocity field without excitations, and the gain in the kinetic energy is mainly due to the superflow of the condensate~\cite{Landau}. The superfluid fraction can be written as~\cite{Burnett}
\beq
f_s= \frac{2m L^2}{\hbar^2{\cal N}_{t}}\lim_{\theta\to 0}\frac{E_\theta-E_0}{ \theta^2}\, ,
\eeq
where $\theta$ is the the phase twist along the lattice and $E_\theta,\,E_0$ are the resulting ground state energies with and without the twist. This difference in energies with and without phase twist can be treated perturbatively for small $\theta$. Within the second order perturbation theory we get the expression for the superfluid fraction as

\begin{widetext}
\beq
f_s=\frac{1}{2}\frac{\sum_i^L (\phi_{i-1}+\phi_{i+1})\phi_i}{{\cal N}_{t}}
-\frac{J}{{\cal N}_{t}} \sum_i^L (\phi_{i-1}-\phi_{i+1})^2\nonumber
\sum_{\nu_i\neq 0}\frac{|\sum_n^\infty \sqrt{n+1} f_{n+1}^{\nu_i}f_{n}^{0_i}|^2}{{\cal E}_{\nu_i}-{\cal E}_{0_i}}\, .
\eeq
\end{widetext}

From this expression it can be seen that the competition between the first and the second order terms give rise to the net superfluid fraction, the role of the second order contribution is to damp and reduce the superfluid component. It is easy to verify that in the homogeneous case, the condensate and superfluid fractions coincide, $f_{s}=f_{c}$.

Atom \textit{number fluctuations} at each lattice site are defined as 
\beq
\Delta n_i=\sqrt { \langle \hat{n}_i ^2 \rangle -\langle \hat{n}_i\rangle ^2}\, ,
\eeq 
and directly characterize the different quantum phases of the system, e.g., in the Mott phase the number fluctuation vanishes $\Delta n_i/n_{i}=0$, whereas in the superfluid phase it becomes large, $\Delta n_i\approx n_i$.

\begin{figure}
\includegraphics[width=1.0\columnwidth]{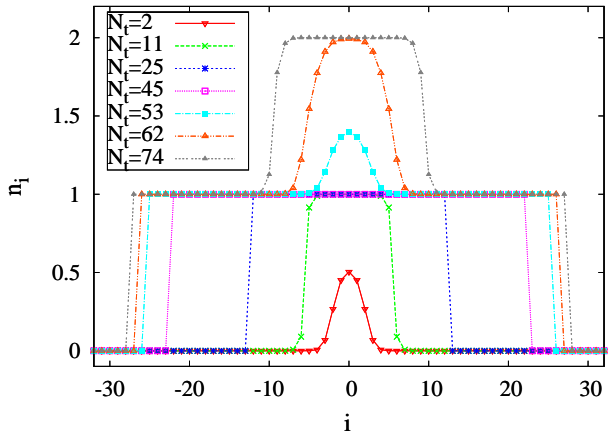}
\caption{(Color online) Atom distribution in the lattice for the trapping strength ${\cal V}_{t}/E_R=0.001$ as the number of atoms is increased from ${\cal N}_{t}=2$ to ${\cal N}_{t}=74$. The length of the lattice is $L=64$ and the height of the primary lattice is $s_1 ^z=16$ corresponding to the ratio $U/J=132$.} 
\label {f1}
\end{figure}
In Fig. \ref{f1} we show a typical ground state density distribution for a trapping strength ${\cal V}_{t}=0.001$ (that corresponds to the longitudinal trapping frequency $\omega_z\approx 2\pi\times 70$Hz) and different values of the total number of atoms, ${\cal N}_{t}$, in the tube. The atoms are delocalized with significant superfluid fraction for a few number of atoms in the tube. As the number of atoms reaches the critical value a flat distribution with Mott-like structure appears near the center. Since the Mott phase is incompressible, further addition of particles uniformly expand the cloud until a non-integer occupancy at the center $\langle \hat n_c\rangle=n+\epsilon$ is energetically favorable. This gives rise to large number fluctuations and superfluid fraction at the center of the trap.

\begin{figure}
\includegraphics[width=1.0\columnwidth]{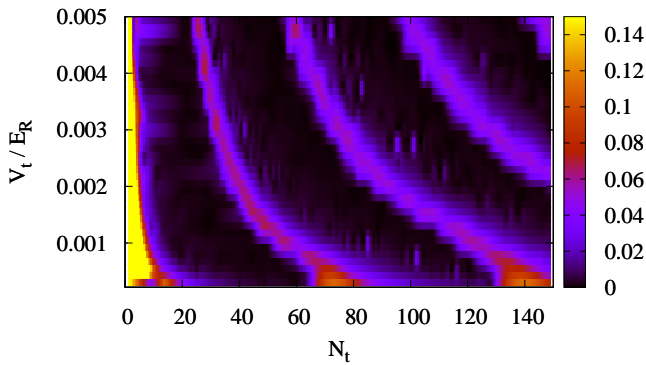}
\caption{(Color online) Density plot of the superfluid fraction in the ${\cal N}_{t}$\,-\,${\cal V}_{t}$ plane for the same parameters of Fig.~\ref{f1}. Each stripe represents the nonvanishing superfluid fraction.}
\label {f2}
\end{figure}
Figure~\ref{f2} depicts the density plot of the superfluid fraction in the ${\cal N}_{t}$\,-\,${\cal V}_{t}$ plane for the parameters discussed in Sec.~IIA. The brighter region represents the nonvanishing superfluid fraction while the darker region corresponds to the insulating state. Each successive superfluid stripe represents the state of the system characterized by the density at the center of the trap with $\langle n_c\rangle=1+\epsilon,\,2+\epsilon,\,3+\epsilon$ etc. These stripes bend towards the smaller number of atoms when the strength of trapping potential is increased. 

\subsection{Size of the cloud and compressibility}

A direct manifestation of the incompressible (compressible) nature of the Mott (superfluid) 
phase can be obtained by measuring the response of the cloud as the trapping frequency is varied. The system is incompressible if the size of the cloud for a fixed atom number remains unchanged. This method has been successfully implemented experimentally to detect the Mott phase in a fermionic system~\cite{Jordens,Schneider}. 

The global compressibility can be related to the size of the cloud $\cal R$ as
\beq
{\kappa}({\cal V}_{t},{\cal N}_{t}\dots) \propto -\frac{\partial {\cal R}}{\partial {\cal V}_{t}}\, ,
\label{eq11}
\eeq
with $\cal R$ being defined as the root-mean-squared (RMS) width 
\beq
\label{eq:rms}
{\cal R}= \sqrt{\sum_j j^2\langle \hat n_j\rangle - \bigg[\sum_j j\langle \hat n_j\rangle\bigg]^2}\, ,
\eeq
in units of lattice constant $d$.
The latter can be measured experimentally with great accuracy by means of in situ imaging technique~\cite{Andrew}.

\begin{figure}
\includegraphics[width=1.0\columnwidth]{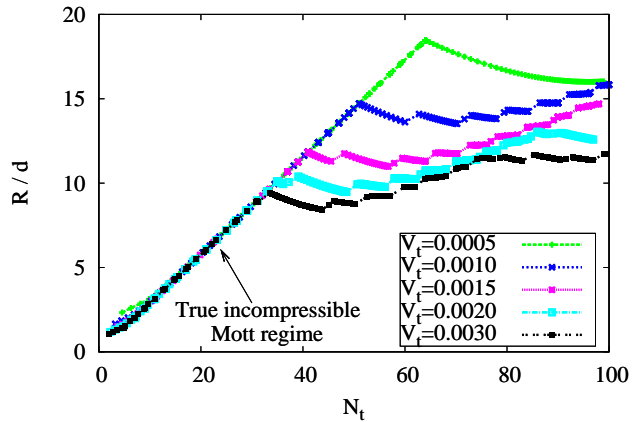}
\caption{(Color online) Cloud size as a function of the total number of atoms for different trapping potentials ${\cal V}_{t}$ in units of $E_R$. The Mott phase is incompressible as the size of the cloud for different trapping potentials for a fixed atom number remains constant. The wide separation between the curves represents the compressible superfluid phase.}
\label {f3}
\end{figure}
In Fig. \ref{f3} we show the RMS size ${\cal R}$ of the atomic cloud as a function of the total number of atoms ${\cal N}_{t}$ for different trapping potentials. In the Mott regime, the size of the  cloud increases linearly with the number of atoms until the potential energy of an atom at the periphery exceeds the onsite interaction energy. A sharp kink in the value of ${\cal R}$ in each curve represents a transition from  occupancy $\langle \hat{n}_{i}\rangle=n$ to $\langle \hat{n}_{i}\rangle=n+\epsilon$ at the center of the trap. This transition occurs earlier for the higher value of the trapping potential.

Once the system exceeds the critical occupancy the size decreases with the addition of atoms since more nearby central sites allow multiple occupancies, and therefore the RMS size decreases according to Eq. (\ref{eq:rms}). A wide separation between these curves for close values of $V_{t}$ indicates that the system is compressible. In the Mott regime, these curves overlap. This is the case of the Mott phase for low filling in Fig.~\ref{f1}, corresponding to the initial linear slope indicated in Fig. \ref{f3}. 

\begin{figure}
\includegraphics[width=1.0\columnwidth]{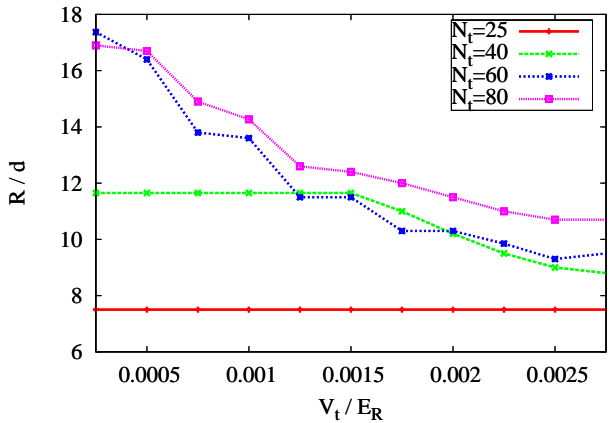}
\caption{(Color online) Cloud size as a function of the trapping potential for different atom numbers ${\cal N}_{t}=25,\,40,\,60$ and $80$.}
\label {f4}
\end{figure}
This can also be seen in Fig.~\ref{f4} where we plot the RMS size as a function of the trapping potential ${\cal V}_{t}$ for different atom numbers ${\cal N}_{t}=25,\,40,\,60$ and $80$. For a small atom number the size virtually remains unaffected, whereas it decreases for larger number of atoms.

The slope of each curve is proportional to the compressibility (see Eq.~(\ref{eq11})). The system is incompressible for  low atom number and for low trapping potentials, when the atom distribution is independent on the trap strength. This situation is the boson counterpart of the fermionic Mott insulator as demonstrated in~\cite{Jordens,Schneider}.

For larger number of atoms the distribution spreads over the outer part of the trap and therefore, an increase of the trapping potential produces a large energy offset with respect to the center of the trap and atoms jump towards the center of the trap resulting a dramatic drop in the cloud size. A similar observation has also been pointed out in Ref.~\cite{Delande} using TEBD calculation. For larger number of atoms far from Mott filling, multiple occupancy smoothly fill the energy gaps and the system always responds to the trap variation resulting a compressible superfluid phase.

The mechanism of supefluid-Mott transition by varying either the number of atoms or the harmonic confinement can also be explained in the phase diagram shown in Fig.~\ref{f2}. One can move along the vertical direction for a fixed number of atoms or along the horizontal direction for a fixed trap strength to see if there is a dramatic increase in the superfluid fraction. Each bump in the superfluid fractions represents the jump in atom number from an integer to non integer value in the center of the trap.

\subsection{\label{sec:level1}Plateaus and emergence of mini gaps}

\begin{figure}
\includegraphics[width=1.0\columnwidth]{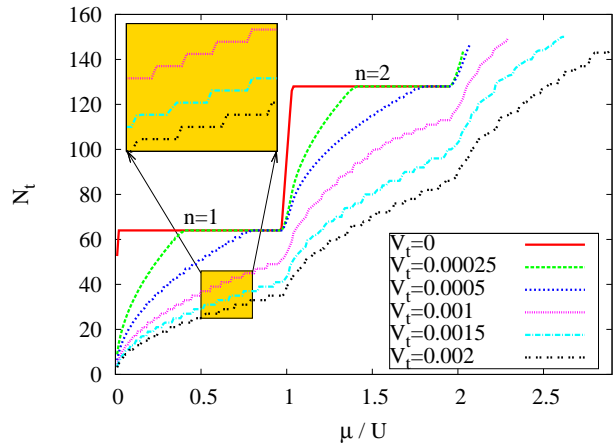}
\caption{(Color online) Variation of the total atom number ${\cal N}_{t}$ as a function of chemical potential $\mu$ for different trapping potentials ${\cal V}_{t}$ ranging from ${\cal V}_{t}=0$ to ${\cal V}_{t}=0.003$ in units of $E_R$. The Mott plateaus
gradually shrinks as ${\cal V}_{t}$ gets stronger and finally disappears when ${\cal V}_{t}>{\cal V}_{t}^c$. For strong confinement the curve develops a series of plateaus (inset) which becomes more pronounced when $\mu/U\rightarrow 1$.}
\label {f5}
\end{figure}

In Fig.~\ref{f5} we show the behavior of the total number of atoms as a function of the chemical potential for different  external trapping potentials for the same parameters considered before. In the Mott regime the curve shows plateaus for integral fillings. In addition to the Mott plateaus the interplay between lattice periodicity and trap inhomogeneity displays a series of additional plateaus in the density versus chemical potential curve (see inset). Such plateaus become more pronounced when the trapping potential is strong. The origin of these discrete jumps in the density is due to the energy offset produced by the harmonic potential at each lattice point. For strong atom-atom interactions the system avoids multiple occupancies and the corresponding energy is minimized by filling the unoccupied lattice site starting successively from the center of the trap analogous to fermions filling in the Fermi sea. As the number of atoms increases the addition of an extra atom to the system already contained ${\cal N}_{t}$ atoms costs energy of the order of $\frac{1}{2}m\omega^2 ({\cal N}_{t}/2)^2 d^2$ which is the energy gap as seen in Fig.~\ref{f5} (inset). As expected this gap becomes more pronounced when the trapping frequency is strong.

In addition to these plateaus Fig. \ref{f5} also shows the disappearance of the Mott plateau as the trapping potential is reached to a critical value ${\cal V}_{t}^c$. For this particular set of parameters the critical trapping potential is ${\cal V}_{t}^c\approx 0.001$. For smaller trap strength, the strong interaction pushes the atoms to the boundary so that the system reaches integer filled homogeneous Mott state. As the trapping potential is increased the true Mott region with $\langle n\rangle=1$ shrinks until double occupancy is more favorable and the atoms start jumping to the central lattice site. Unlike in the homogeneous system, the nonzero average slope in the ${\cal N}_{t}$ versus $\mu$ plot for each value of the trapping potential gives a small but finite compressibility and hence the trapped system is, in principle, always compressible. Also note that the compressibility $\kappa$ tends to vanish when $\mu/U\rightarrow 1$ for each value of ${\cal V}_{t}> {\cal V}_{t}^c$ since $\kappa=\partial n/\partial \mu$.

\section{\label{sec:level1}Effects of secondary lattice: Disordered case}

In this section we consider the situation when an additional secondary lattice, incommensurate to the main one is introduced resulting a quasiperiodic system as discussed in Sec.~II. Quasiperiodic disorder can be controlled by varying the intensity of the secondary laser  as the strength, $\Delta$, is proportional to the height of the secondary lattice $s_2^z$ (see Eq. (\ref{eq:delta})). The lattice incommensurate parameter gives an additional characteristic periodicity of $1/(1-\alpha)$. 

In the following we will consider $\alpha=(\sqrt 5-1)/2$, the inverse of the golden mean~\cite{Aulbach}.
In a non-interacting system the threshold for the extended to the localized single particle state occurs at $\Delta/J=2$~\cite{Aubry}. In the strongly interacting regime, $J\ll U$, the interaction is the only dominant energy scale. Therefore, the localization may occur as soon as the disorder is switched on~\cite{Fisher}. 

\subsection{Size of the cloud and compressibility}

\begin{figure}
\includegraphics[width=1.0\columnwidth]{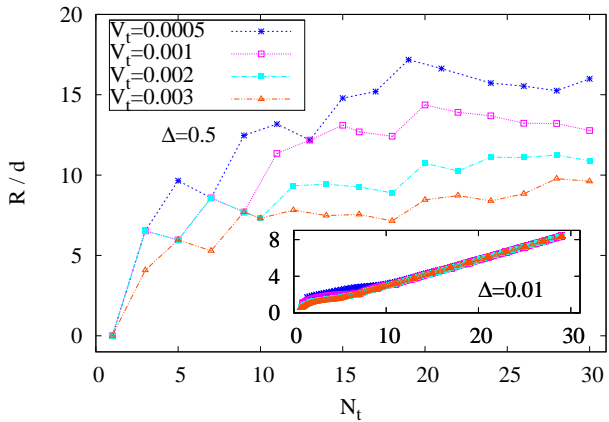}
\caption{(Color online) Cloud size as a function of the total number of atoms for different trapping potentials for the disorder strength $\Delta=0.5$ and $\Delta=0.01 (\approx 2J)$ (shown in the inset). Both $\Delta$ and ${\cal V}_{t}$ are expressed in terms of the recoil energy $E_R$.}
\label {f6}
\end{figure}

The effect of trapping potential on the size of the cloud in the presence of the secondary lattice is demonstrated in Fig.~\ref{f6}, where we plot the RMS size $\cal R$ as a function of the total atom number ${\cal N}_{t}$, for two values of the disorder strength namely $\Delta=0.5\, (\gg J)$, which is the critical value of the disorder strength for the disappearance of the Mott lobe~\cite{Fisher}, and $\Delta=0.01(\approx 2J)$ for which the localization-delocalization transition in the noninteracting system occurs~\cite{Aubry}. The rest of the parameters are the same as in the pure case. The robust nature of the Mott insulating state for small $\Delta=0.01$ (shown in inset) and a wide separation in $\cal R$ for large $\Delta=0.5$, as the trapping strength is varied, manifestly indicate that the former state is incompressible while the latter state is compressible. It is noted that there are certain points for $\Delta=0.5$ at which the curves for different values of the trapping potentials tend to cling together when the number of atoms are small. This behavior may be accounted due to the presence of additional plateaus as seen in Fig.~\ref{f5}.

\begin{figure}
\includegraphics[width=1.0\columnwidth]{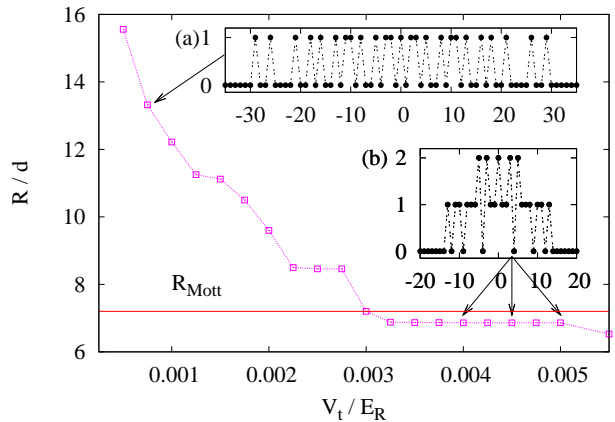}
\caption{(Color online) Cloud size as a function of the trapping potential for the total atom number in the tube ${\cal N}_{t}=25$. Also shown (in the insets) are the atom densities in the weak (a) and strong (b) confinements.}
\label {f7}
\end{figure}
In Fig.~\ref{f7} we plot the cloud size ${\cal R}$ as a function of trapping potential for a fixed atom number ${\cal N}_{t}=25$, corresponding to the Mott state in the pure system. For small trapping potential the size $\cal R$ decreases monotonically. We also show the typical state of the system for the two values of the trapping potentials (in the insets). The atom density is inhomogeneously distributed due to the quasiperiodicity with many unoccupied lattice sites in between. As the trapping potential increases more and more atoms from the edges displace towards the center of the trap or near the quasiperiodic minima and thus reducing the size of the cloud. This decrease in ${\cal R}$ along with the vanishing superfluid fraction may hint that the system has entered into the compressible (quasi) Bose-glass phase. This prescription works fine as long as the trapping potential is weak.

As the trapping potential gets stronger, atoms start accumulating near the center of the trap and the size $\cal R$ develops intervening plateaus, indicating incompressible nature of the system. These plateaus becomes more pronounced when the trapping potential is strong. From the inset (b) of Fig.~\ref{f7} we see that the state of the system does not change over a wide range of trapping potentials. This incompressible state in the presence of disorder differs from the Mott phase particularly in its density distribution. The atom distributions are still inhomogeneous with many intervening unoccupied along with the doubly occupied sites. These insulating states are the characteristics of the quasiperiodic potential in the presence of trapping potential and resemble to the so-called Mott-glass phase as reported in Ref.~\cite{Sengupta}. 

\section{\label{sec:level1}Concluding remarks} 

In conclusion we have studied the possibility of detecting quantum phases of strongly interacting bosons in a one dimensional optical lattice in the presence of trapping and quasiperiodic potentials. In general the trapping potential may obscure the superfluid-Mott transition and a direct mapping of the phase diagram from the inhomogeneous to the homogeneous system can be challenging. In particular implementing the idea of trap-squeezing to detect the quantum phase poses challenges due to the local nature of the system~\cite{RoscildeNJP}. 

In this paper we have studied the effects of trapping potential in a strongly interacting bosonic system by varying the number of atoms. In a pure system a true incompressible Mott state can be observed for small number of atoms over a range of trapping potential. In parallel to the recent experiments on detecting the incompressibility of fermions~\cite{Jordens,Schneider}, the robust nature of the Mott state is verified by varying the trapping potentials as the cloud size remains unaffected. However, the introduction of the disorder induces trap-dependent cloud size. The vanishing superfluid fraction along with the trap-dependent cloud size may hint that the system is in Bose-glass phase.

In addition to the compressible phase, the strong trapping confinement may induce a Mott-like incompressible states in the presence of disorder. These insulating states differ from Mott phase in their density distribution. The strongly interacting system in the presence of disorder interpolates between two insulating states namely incompressible Mott-like states and compressible Bose-glass as the trapping potential is continuously varied.

\acknowledgments{This work was supported by the EU Contract EU STREP NAMEQUAM. We thank people at LENS QDG group and T. Roscilde for fruitful discussions. We also thank J. Javanainen for reading the manuscript carefully.}

\end{document}